\input psfig.sty
\documentstyle[12pt]{article}
\def\beq{\begin{equation}}
\def\eeq{\end{equation}}
\def\beqn{\begin{eqnarray}}
\def\eeqn{\end{eqnarray}}
\def\twiddle{\lower.9ex\rlap{$\kern-.1em\scriptstyle\sim$}}
\def\bigtwiddle{\lower1.ex\rlap{$\sim$}}
\def\gtwid{\mathrel{\raise.3ex\hbox{$>$\kern-1em\lower1ex\hbox{
$\scriptstyle\sim$}}}}
\def\ltwid{\mathrel{\raise.3ex\hbox{$<$\kern-.75em\lower1ex\hbox{
$\sim$}}}}

\def\neweq{\setcounter{equation}{0}}

\begin{document}
\title{DARK MATTER AXIONS '96}
\author{Pierre Sikivie\\
{\it Department of Physics}\\
{\it University of Florida}\\
{\it Gainesville, FL  32611}}
\date{}
\maketitle

\vspace{.2in}
\section*{Contents:}

\begin{enumerate}
\item The strong CP problem
\item Dark matter axions
\item The cavity detector of galactic halo axions
\item The phase space structure of cold dark matter halos
\end{enumerate}

\section{The strong CP problem}

The axion was postulated nearly two decades ago \cite{arev} to explain 
why the strong interactions conserve $P$ and $CP$ in spite of the fact 
that the weak interactions violate those symmetries.  Consider the 
Lagrangian of QCD:
\beq
{\cal L}_{QCD} = -{1\over 4} G^a_{\mu\nu} G^{a\mu\nu} +\sum_{j=1}^n
\left[ \overline q_j \gamma^\mu i D_\mu q_j - (m_j q^+_{Lj} q_{Rj}
+ \hbox{h.c.})\right]
+ {\theta g^2\over 32\pi^2} G^a_{\mu\nu} \tilde G^{a\mu\nu} \,\,\,  .
\eeq
The last term is a 4-divergence and hence does not contribute in 
perturbation theory.  That term does however contribute through 
non-perturbative effects \cite{thft} associated with QCD instantons 
\cite{inst}.  Such effects can make the physics of QCD depend upon the 
value of $\theta$. Using the Adler-Bell-Jackiw anomaly \cite{abj}, one 
can show that $\theta$ dependence must be there if none of the current 
quark masses vanishes.  (If this $\theta$ dependence were absent,
QCD would have a $U_A(1)$ symmetry and would predict the mass of 
the $\eta'$ pseudo-scalar meson to be less than 
$\sqrt{3} m_\pi \approx 240$~MeV \cite{SW},contrary to observation.)  
One can further show that QCD depends upon $\theta$ only through the 
combination of parameters:
\beq
\overline \theta =\theta -\hbox{arg} (m_1, m_2, \ldots m_n)
\eeq
If $\overline \theta \neq 0$, QCD violates $P$ and $CP$.  The absence of
$P$ and $CP$ violations in the strong interactions therefore places an
upper limit upon $\overline\theta$.  The best constraint follows from 
the experimental bound \cite{ned} on the neutron electric dipole moment 
which yields: $\overline\theta < 10^{-9}$.

The question then arises:  why is $\overline\theta$ so small?  In the 
standard model of particle interactions, the quark masses originate in 
the electroweak sector of the theory.  This sector must violate $P$ and 
$CP$ to produce the correct weak interaction phenomenology.  There is 
no reason in the standard model to expect the overall phase of the quark 
mass matrix to exactly match the value of $\theta$ from the QCD sector so 
that $\overline \theta < 10^{-9}$. In particular, if $CP$ violation is 
introduced in the manner of Kobayashi and Maskawa \cite{KM}, the Yukawa 
couplings that give masses to the quarks are arbitrary complex numbers 
and hence arg~det~$m_q$ and $\overline\theta$ have no reason to take on 
any special value at all.

The problem why $\overline\theta < 10^{-9}$ is usually referred to as the 
``strong $CP$ problem''.  The existence of an axion would solve this 
problem.  But, before we talk about axion physics, let's mention that there 
are other solutions.  Setting $m_u = 0$ removes the $\theta$-dependence 
of QCD and hence the strong $CP$ problem as well.  However, $m_u = 0$ 
may cause problems with the successful current algebra relations among 
pseudo-scalar meson masses.  I refer the reader to refs.\cite{NS,Leut} 
for recent discussions of the issues involved.  Another type of solution 
involves the assumption that $CP$ and/or $P$ is spontaneously broken but 
is otherwise a good symmetry.  In this case, $\overline\theta$ is calculable 
and may be arranged to be small \cite {CPsb}. Finally, let's emphasize that 
the strong $CP$ problem need not be solved in the low energy theory.  
Indeed, as Ellis and Gaillard \cite{EG} pointed out, if in the standard 
model $\overline\theta=0$ near the Planck scale then 
$\overline\theta \ll 10^{-9}$ at the QCD scale.

Peccei and Quinn \cite{PQ} proposed to solve the strong $CP$ problem by 
postulating the existence of a global $U_{PQ}(1)$ quasi-symmetry.  To do 
its job, $U_{PQ}(1)$ must be a symmetry of the theory at the classical 
(i.e., at the Lagrangian) level, it must be broken explicitly by those 
non-perturbative effects that make the physics of QCD depend upon $\theta$, 
and finally it must be spontaneously broken.  The axion \cite{WW} is the 
pseudo-Nambu-Goldstone boson associated with the spontaneous breakdown 
of $U_{PQ}(1)$. One can show that, if a $U_{PQ}(1)$ quasi-symmetry is 
present, then
\beq
\overline\theta = \theta - arg (m_1 \ldots m_n) - {a(x)\over f_a}\, ,
\eeq
where $a(x)$ is the axion field and $f_a$, called the axion decay 
constant, is of order the vacuum expectation value (VEV) which spontaneously
breaks $U_{PQ}(1)$. \ It can further be shown \cite{VW} that the 
non-perturbative effects that make QCD depend upon $\overline\theta$ 
produce an effective potential $V(\overline\theta)$ whose minimum is 
at $\overline\theta =0$.\ \ Thus, by postulating an axion, $\overline\theta$ 
is allowed to relax to zero dynamically and the strong $CP$ problem is solved.

The properties of the axion can be derived using the methods of current
algebra \cite{curr}.  The axion mass is given in terms of $f_a$ by
\beq 
m_a\simeq 0.6~eV~{10^7 GeV\over f_a}\, .
\eeq
All the axion couplings are inversely proportional to $f_a$.  For example,
the axion coupling to two photons is:
\beq
{\cal L}_{arr} = -g_\gamma {\alpha\over \pi} {a(x)\over f_a}
\vec E \cdot\vec B
\eeq
where $\vec E$ and $\vec B$ are the electric and magnetic fields, 
$\alpha$ is the fine structure constant, and $g_\gamma$ is a model-dependent 
coefficient of order one.  $g_\gamma=0.36$ in the DFSZ model \cite{DFSZ}
whereas $g_\gamma=-0.97$ in the KSVZ model \cite{KSVZ}.  A priori the 
value of $f_a$, and hence that of $m_a$, is arbitrary.  However, searches 
for the axion in high energy and nuclear physics experiments combined with
astrophysical constraints, the latter derived by considering the effect
of the axion upon the lifetimes of red giants and SN1987a, rule out 
$m_a \gtwid 10^{-3}$~eV \cite{arev}. \ In addition, as will be discussed 
in section~II, cosmology places a lower limit on $m_a$ of order
$10^{-6}$~eV by requiring that axions do not overclose the universe.
\neweq

\section{Dark matter axions}
For small masses, axion production in the early universe is dominated
by a novel mechanism \cite{ac}.  The crucial point is that the 
non-perturbative QCD effects that produce the effective potential 
$V(\overline\theta)$ are strongly suppressed at temperatures high 
compared to $\Lambda_{QCD}$ \cite{GPY}. At these high temperatures, 
the axion is massless and all values of $\langle a(x)\rangle$ are 
equally likely.  At $T\simeq 1$~GeV, the potential $V$ turns on and 
the axion field starts to oscillate about a $CP$ conserving minimum 
of $V$. These oscillations do not dissipate into other forms of energy 
because, in the relevant mass range, the axion is too weakly coupled for 
that to happen.  The oscillations of the axion field may be described as 
a fluid of axions.  The typical momentum of the axions in the fluid is 
the inverse of the correlation length of the axion field.  Because that 
correlation length is of order the horizon, we have 
$p_a \sim (10^{-6}$ sec)$^{-1} \sim 10^{-9}$~eV at $T\simeq 1$~GeV, and 
$p_a \sim R^{-1}$ afterwards.  $R$ is the cosmological scale factor here.  
Thus the axion fluid is very cold compared to the ambient temperature.

Let me briefly indicate how the present cosmological energy density of this
axion fluid is estimated.  Let $\varphi(x)$ be the complex scalar field 
whose VEV $v$ spontaneously breaks $U_{PQ}(1)$. \ At extremely high 
temperatures, the $U_{PQ}(1)$ symmetry is restored.  It becomes spontaneously 
broken when the temperature drops below a critical value $T_{PQ}$ of order 
$v$. Below $T_{PQ}$, the axion field $a(x)$ appears as the phase of the 
$VEV$ of $\varphi$:  $\langle \varphi(x) \rangle = ve^{ia(x)/v}$.  
We must now distinguish two cases.  Either inflation occurs with reheat 
temperature below $T_{PQ}$, or not (i.e., inflation does not occur or it 
occurs with reheat temperature above $T_{PQ}$).  In the first case, inflation 
homogenizes the axion field and there is only one contribution to the axion 
cosmological energy density, the contribution from so-called ``vacuum 
misalignment''.  In the second case, there are additional contributions 
from axion string and axion domain wall decay.  Only the contribution from 
vacuum misalignment is  discussed in any detail here.

When the axion mass turns on near the QCD phase transition, the axion
field starts to oscillate about one of the $CP$ conserving minima of
the effective potential.  The oscillation begins approximately at
cosmological time $t_1$ such that $t_1m_a (T(t_1)) = 0(1)$, where 
$m_a(T)$ is the temperature dependent axion mass.  Soon after $t_1$,
the axion mass changes sufficiently slowly that the total number of
axions in the oscillations of the axion field is an adiabatic
invariant.  $T_1 \equiv T(t_1)$ has been estimated to be of order 
1~GeV.\ \ The number density of axions at time $t_1$ is
\beq
n_a(t_1)\simeq {1\over 2} m_a(t_1) \langle a^2(t_1)\rangle \simeq
\pi f_a^2 {1\over t_1}
\eeq
where $f_a = {v\over N}$ is the axion decay constant introduced 
earlier.  $N$ is an integer which expresses the color anomaly of
$U_{PQ}(1)$. \ $N$ also equals the number of $CP$ conserving vacua \cite{adw}
at the bottom of the 'Mexican hat' potential, i.e., in the interval 
$0\leq {a\over v} < 2\pi$.  In Eq.~(2.1), we used the fact that the 
axion field $a(x)$ is approximately homogeneous on the horizon scale 
$t_1$.  Wiggles in $a(x)$ which entered the horizon long before $t_1$ 
have been red-shifted away \cite{Vil}.  We also used the fact 
that the initial departure of $a(x)$ from the nearest minimum is of 
order ${v\over N} = f_a$. \ The axions of Eq.~(2.1) are decoupled and
non-relativistic.  Assuming that the ratio of the axion number 
density to the entropy density is constant from time $t_1$ till
today, one finds \cite{ac}
\beq
\Omega_a = \left({0.6~10^{-5}\hbox{\ eV}\over m_a}\right)^{7\over 6}
\left({200\hbox{\ MeV}\over \Lambda_{QCD}}\right)^{3\over 4}
\left({75\hbox{\ km/s}\cdot \hbox{Mpc}\over H_0}\right)^2
\eeq
for the ratio of the axion energy density to the critical density
for closing the universe.  $H_0$ is the present Hubble rate.
Eq.~(2.2) implies the bound $m_a \gtwid 10^{-6}$~eV. 

However, it should be emphasized that there are many sources of 
uncertainty in the estimate of Eq.~(2.2). \ The axion energy
density may be diluted by the entropy release from heavy particles
which decouple before the QCD epoch but decay afterwards \cite{ST}, 
or by the entropy release associated with a first order QCD phase 
transition.  On the other hand, if the QCD phase transition is
first order \cite{pt}, an abrupt change of the axion mass at the 
transition may increase $\Omega_a$.  If inflation occurs with reheat 
temperature less than $T_{PQ}$, there may be an accidental
suppression of $\Omega_a$ because the homogenized axion field
happens to lie close to a $CP$ conserving minimum.  Because the
RHS of Eq.~(2.2) is multiplied in this case by a factor of order the 
square of the initial vacuum misalignment angle ${a(t_1)\over v}N$
which is randomly chosen between $-\pi$ and $+\pi$, the probability 
that $\Omega_a$ is suppressed by a factor $x$ is of order $\sqrt{x}$.  
This rule cannot be extended to arbitrarily small $x$ however because 
quantum mechanical fluctuations in the axion field during the epoch of 
inflation do not allow the suppression to be perfect \cite{inflax}.  
If inflation occurs with reheating temperature larger than $T_{PQ}$ 
or if there is no inflation, there are contributions to $\Omega_a$ 
from axion string \cite{rd} and axion domain wall decay in addition 
to the contribution, Eq.~(2.2), from vacuum misalignment.  The author and 
his collaborators \cite{ssa} have estimated each of these additional 
contributions to be of the same order of magnitude as that from vacuum 
misalignment.  Others \cite{rd,osa} have estimated that the contribution from 
axion string decay dominates over that from vacuum misalignment by a 
factor 100 or a factor 10.

The axions produced when the axion mass turns on during the 
QCD phase transition are cold dark matter (CDM) because the
axions are non-relativistic from the moment of their first
appearance at 1~GeV temperature.  Studies of large scale
structure formation support the view that the dominant fraction 
of dark matter is CDM \cite{MST}.  Moreover. any form of CDM necessarily
contributes to galactic halos by falling into the gravitational
wells of galaxies.  Hence, there is excellent motivation to look
for CDM candidates as constituent particles of our galactic
halo, even after some fraction of our halo has been demonstrated
to be in MACHOs \cite{MACHO} or some other form.

Finally, let's mention that there is a particular kind of clumpiness 
\cite{amc} which affects axion dark matter if there is no inflation after 
the Peccei-Quinn phase transition.  This is due to the fact that the
dark matter axions are inhomogeneous with $\delta \rho / \rho \sim 1$ over 
the horizon scale at temperature $T_1 \simeq$ 1 GeV, when they 
are produced at the start of the QCD phase-transition, combined 
with the fact that their velocities are so small that they do not 
erase these inhomogeneities by free-streaming before the time $t_{eq}$
of equality between the matter and radiation energy densities when 
matter perturbations can start to grow.  These particular inhomogeneities 
in the axion dark matter are immediately in the non-linear regime after 
time $t_{eq}$ and thus form clumps, called `axion mini-clusters' 
\cite{amc}.  These have mass $M_{mc} \simeq 10^{-13} M_\odot$ and 
size $l_{mc} \simeq 10^{12}$ cm.
\neweq

\section{The cavity detector of galactic halo axions}
Axions can be detected by stimulating their conversion to photons in a 
strong magnetic field \cite{ps}.  The relevant coupling is given in 
Eq.~(1.5).  In particular, an electromagnetic cavity permeated by a 
strong static magnetic field can be used to detect galactic halo axions.  
The latter have velocities $\beta$ of order $10^{-3}$ and hence their 
energies $E_a=m_a+{1\over 2} m_a\beta^2$ have a spread of order 
$10^{-6}$ above the axion mass.  When the frequency $\omega=2\pi f$ of a
cavity mode equals $m_a$, galactic halo axions convert resonantly
into quanta of excitation (photons) of that cavity mode.  The
power from axion $\to$ photon conversion on resonance is found
to be \cite{ps,kal}:
\beqn
P &=& \left ({\alpha\over\pi} {g_\gamma\over f_a}\right )^2 V\, B_0^2 
\rho_a C {1\over m_a} \hbox{Min}(Q_L,Q_a)\nonumber\\
&=& 0.5\; 10^{-26} \hbox{Watt}\left( {V\over 500\hbox{\ liter}}
\right) \left({B_0\over 7\hbox{\ Tesla}}\right)^2 C\left({g_\gamma\over
0.36}\right)^2 \nonumber\\
&\cdot& \left({\rho_a\over {1\over 2} \cdot 10^{-24}
{{\scriptstyle g_r}\over \hbox{\scriptsize cm}^3}}\right) 
\left({m_a\over 2\pi (\hbox{GHz})}\right) \hbox{Min}(Q_L,Q_a)
\eeqn
where $V$ is the volume of the cavity, $B_0$ is the magnetic field 
strength, $Q_L$ is its loaded quality factor, $Q_a=10^6$ is the 
`quality factor' of the galactic halo axion signal (i.e. the ratio of 
their energy to their energy spread), $\rho_a$ is the density of galactic
halo axions on Earth, and $C$ is a mode dependent form factor
given by
\beq
C = {\left| \int_V d^3 x \vec E_\omega \cdot \vec B_0\right|^2
\over B_0^2 V \int_V d^3x \epsilon |\vec E_\omega|^2}  \, 
\eeq
where $\vec B_0(\vec x)$ is the static magnetic field,
$\vec E_\omega(\vec x) e^{i\omega t}$ is the oscillating electric
field and $\epsilon$ is the dielectric constant.

Because the axion mass is only known in order of magnitude at
best, the cavity must be tunable and a large range of frequencies
must be explored seeking a signal.  The cavity can be tuned by
moving a dielectric rod or metal post inside it.  Using Eq.~(3.1), one 
finds that to perform a search with signal to noise ratio $s/n$, the 
scanning rate is:
\beqn
{df\over dt} &=& {12 \hbox{GHz}\over \hbox{year}} \left({4n\over s}
\right)^2 \left({V\over 500\hbox{\ liter}}\right)^2
\left( {B_0\over 7\hbox{\ Tesla}}\right)^4 C^2\left({g_\gamma\over
0.36}\right)^4  \nonumber\\
&\cdot &\left({\rho_a\over {1\over 2}\cdot 10^{-24} 
{{\scriptstyle gr}\over \hbox{\scriptsize cm}^3}}\right)^2 
\left({3K\over T_n}\right)^2 \left(
{f\over \hbox{GHz}}\right)^2 {Q_L\over Q_a} \,\,\,  ,
\eeqn
where $T_n$ is the sum of the physical temperature of the cavity plus 
the noise temperature of the microwave receiver that detects the photons 
from $a \to \gamma$ conversion.  Eq.~(3.3) assumes that $Q_L < Q_a$ and 
that some strategies have been followed which optimize the search rate.
The best quality factors attainable at present, using oxygen
free copper, are of order $10^5$ in the GHz range. To make 
the cavity of superconducting material is probably not useful 
since it is permeated by a strong magnetic field in the experiment.

\begin{figure}
\psfig{file=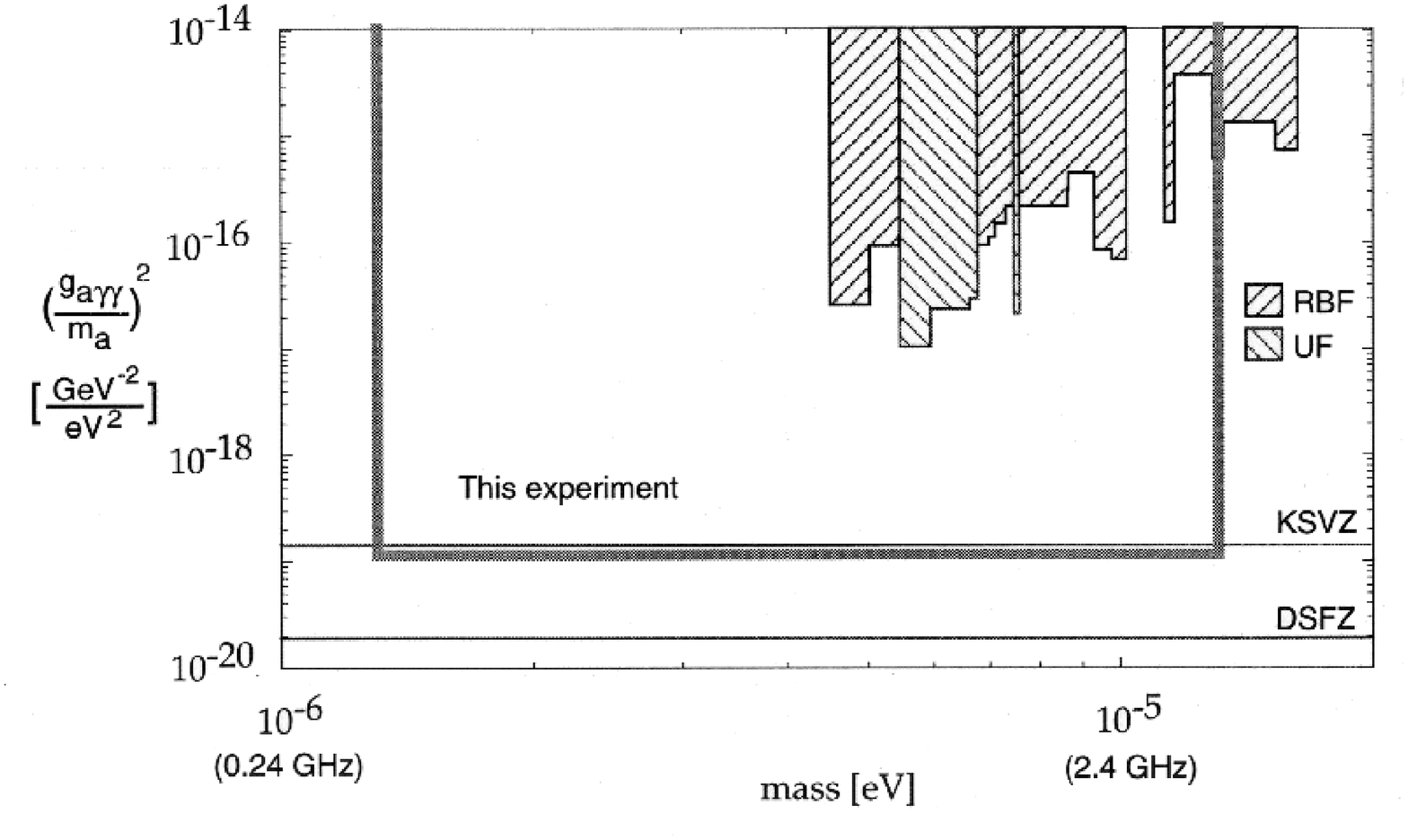,width=6in}
\caption{Regions in mass - (coupling/mass)$^2$ space which have been ruled 
out by the RBF and UF experiments (hatched) and which the LLNL 
experiment expects being able to rule out (shaded border).  The relation 
between coupling and mass in the DFSZ and KSVZ models is also shown.}
\end{figure}

Eq. (3.3) shows that a galactic halo search at the required
sensitivity is feasible with presently available technology,
provided the form factor $C$ can be kept at values of order one
for a wide range of frequencies.  For a cylindrical cavity and a 
longitudinal magnetic field, $C=0.69$ for the lowest TM mode.
The form factors of the other modes are much smaller.  The 
resonant frequency of the lowest TM mode of a cylindrical cavity
is $f$=115 MHz $\left( {1m\over R}\right)$ where $R$ is the radius
of the cavity.  Since $10^{-6}\hbox{\ eV} = 2\pi$ (242 MHz), a
large cylindrical cavity is convenient for searching the low
frequency end of the range of interest.  To extend the search
to high frequencies without sacrifice in volume, one may
power-combine many identical cavities which fill up the
available volume inside a magnet's bore \cite{rsci,hag}.  This 
method allows one to maintain $C=0(1)$ at high frequencies, albeit 
at the cost of increasing engineering complexity as the frequency, 
and hence the number of cavities, is increased.

Pilot experiments were carried out at Brookhaven National
Laboratory \cite{RBF} and at the University of Florida \cite{UF}.  
The (magnetic field)$^2 \times$ volume provided by the magnets used 
in these experiments were relatively low:  $B_0^2V=0.36T^2m^3$ and 
$0.45 T^2 m^3$ respectively for Brookhaven (RBF) and Florida (UF).  
Fig.~I shows the limits that these experiments placed on the square 
of the coupling 
$g_{a\gamma\gamma} = {\alpha\over\pi} {g_\gamma\over f_a}$ as a 
function of the axion mass $m_a$ assuming that the local density of 
galactic halo axions is
$\rho_a ={1\over 2}\cdot 10^{-24} gr/\hbox{cm}^3$.  The figure
also shows $g_{a\gamma\gamma}$ as a function of $m_a$ in the KSVZ and
DFSZ models.

Second generation experiments are presently under way at Lawrence 
Livermore National Laboratory (LLNL) \cite{LLNL} and at Kyoto
University \cite{Kyoto}.  The LLNL experiment uses a much larger 
magnet ($B_0^2 V = 12 T^2m^3$) than the pilot experiments.  It
also improves the noise temperature ($T_n = 3K$ vs. $T_n=5K$ for 
the Florida experiment) although it uses the same microwave receiver 
technology (HEMT amplifiers).  The LLNL experiment is also the first 
to use multiple cavity arrays to expand widely the mass range searched.  
It will cover $1.3 < m_a < 13 \mu$eV at a level of sensitivity sufficient to 
discover KSVZ axions if they are the constituents of our galactic 
halo; see Fig.~I.  \ The LLNL experiment started taking data in 
Feb.~'96 and will run for about three years to cover the stated range.

The Kyoto experiment has a magnet of size similar to that of the pilot 
experiments but uses a beam of Rydberg atoms to count the photons from 
$a\to \gamma$ conversion.  The $a\to \gamma$ conversion part is the 
same as in the other experiments.  Single photon counting constitutes 
a dramatic improvement in microwave detection sensitivity.  With HEMT 
amplifiers one needs to have thousands of $a\to \gamma$ conversions per 
second and integrate for about 100~sec to find a signal in the noise.  
With single photon counting, a few $a\to \gamma$ conversions suffice in 
principle.  To build a beam of Rydberg atoms capable of single photon 
counting is a considerable achievement.  In addition, the cavity will 
be cooled by a dilution refrigerator down to a temperature 
($\sim 10$~mK) where the thermal photon background is negligible.  The 
Kyoto experiment will first search near $m_a=10^{-5}$~eV. \ \ Its 
projected sensitivity is sufficient to discover DFSZ axions even if
their local density is only $1\over 5$ of the local halo density.
\neweq

\section{The phase space structure of cold dark matter halos}
If a signal is found in the cavity detector of galactic halo 
axions, it will be possible to measure their energy spectrum 
with great precision and resolution because all the time 
previously used in searching for the signal can now be used to 
accumulate data. Hence there is good motivation to ask what can 
be learned about our galaxy from analyzing such a signal.

In many past discussions of dark matter detection on Earth, 
it has been assumed that the dark matter particles have an 
isothermal distribution. Thermalization has been argued to be
the result of a period of "violent relaxation'' following the 
collapse of the protogalaxy. If it is strictly true that the 
velocity distribution of dark matter particles is isothermal, 
which seems to be a strong assumption, then the only 
information that can be gained from its observation is the 
corresponding virial velocity and our own velocity relative to 
its standard of rest. If, on the other hand, thermalization 
is incomplete, a signal in a dark matter detector may yield 
additional information.

J.R. lpser and I discussed \cite{jri} the extent to which the 
phase-space distribution of cold dark matter particles is thermalized 
in a galactic halo and concluded that there are substantial deviations
from a thermal distribution in that the highest energy particles 
have discrete values of velocity.  There is one velocity peak on 
Earth due to dark matter particles falling onto the galaxy for the
first time, one peak due to particles falling out of the galaxy 
for the first time, one peak due to particles falling into the 
galaxy for the second time, etc.  The peaks due to particles that 
have fallen in and out of the galaxy a large number of times in the 
past are washed out because of scattering in the gravitational 
wells of stars, globular clusters and large molecular clouds. But 
the peaks due to particles which have fallen in and out of the 
galaxy only a small number of times in the past are not washed out.

\begin{figure}
\psfig{file=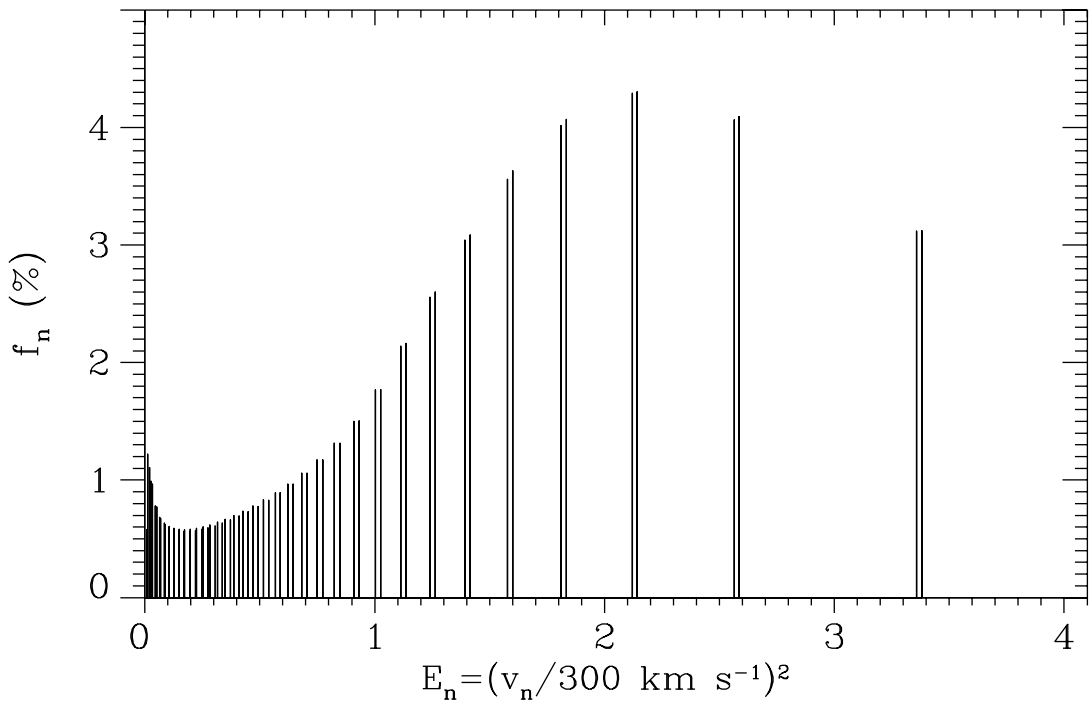,width=6in}
\caption{The spectrum of velocity peaks in a typical case ($\epsilon = 0.2,
h = 0.7$ and $\bar{j} = 0.2$) studied in ref. \cite{stw}.  $f_n$ and 
$E_n$ are defined in the text.}
\end{figure}

I. Tkachev, Y. Wang and I have used the secondary infall model of
galactic halo formation to estimate the local densities and the
velocity magnitudes of the dark matter particles in the velocity
peaks \cite{stw}.  We generalized the existing version of that model 
to take account of the angular momentum of the dark matter particles.  
In the absence of angular momentum, the model produces flat rotation
curves for a large range of values of a parameter $\epsilon$ whose 
average value may be inferred from the spectrum of primordial density 
perturbations.  We find that the presence of angular momentum produces 
an effective core radius, i.e., it makes the contribution of the halo 
to the rotation curve go to zero at zero radius.  The model provides a 
detailed description of the large scale properties of galactic
halos including their density profiles, their extent and their
total mass.  Fig.~II shows the predictions of the model for the
average density fractions $f_n=\rho_n/\rho$ and the
kinetic energies per unit mass $E_n$ of the particles in the
highest energy peaks for the case where $\epsilon=0.2$,
$H_0=70$~km/sec$\cdot$Mpc and the average amount of angular
momentum, in the dimensionless units defined in ref. \cite{stw}, is 
$\overline j=0.2$.  The density fractions $f_n$ are averages over
all locations at the same distance (8.5~kpc) from the galactic 
center as we are.  The $E_n$ are measured in a frame of reference
which is not co-rotating with the disk.

\section*{Acknowledgements:}

I thank the Aspen Center for Physics for its hospitality while writing
up this lecture.  This work is supported in part by the US Department 
of Energy under grant No. DEFG05-86ER40272.

\end{document}